\documentstyle[aps,twocolumn,amsfonts]{revtex}

\newcommand{\li}{\mathop{\rm Li}\nolimits}

\newcommand{\ba}{\begin{eqnarray}}
\newcommand{\ea}{\end{eqnarray}}
 \reversemarginpar
\begin{document}
\title{Nonrelativistic Fermions in Magnetic Fields: a Quantum
Field Theory Approach}
\author{O. Espinosa$^{1}$\thanks{%
E-mail: espinosa@fis.utfsm.cl}, J. Gamboa$^{2}$\thanks{%
E-mail: jgamboa@lauca.usach.cl}, S. Lepe$^{2,3}$\thanks{%
E-mail: slepe@lauca.usach.cl} and F. M\'endez$^{2}$\thanks{
E-mail: fmendez@lauca.usach.cl}}
\address{$^1$Departamento de F\'{\i}sica, Universidad T\'{e}cnica Federico Santa
Mar\'{\i}a, Casilla 110-V, Valpara\'{\i}so, Chile. \\
$^2$Departamento de F\'{\i}sica, Universidad de Santiago de Chile, Casilla
307, Santiago 2, Chile. \\
$^3$Instituto de F\'{\i}sica, Universidad Cat\'{o}lica de Valpara\'{\i}so,
Casilla 4059, Valpara\'{\i}so, Chile.}
\maketitle
\begin{abstract}
The statistical mechanics of nonrelativistic fermions in a
constant magnetic field is considered from the quantum field
theory point of view. The fermionic determinant is computed using
a general procedure that contains all possible regularizations.
The nonrelativistic grand-potential can be expressed in terms
polylogarithm functions, whereas the partition function in 2+1
dimensions and vanishing chemical potential can be compactly
written in terms of the Dedekind eta function. The strong and weak
magnetic fields limits are easily studied in the latter case by
using the duality properties of the Dedekind function.
\end{abstract}
\pacs{05.m0.-d, 05..30.-Fk}
\preprint{USACH-FM-00-03}
\preprint{USM-TH-xx}
\narrowtext
\section{Introduction}
The motion of fermions in magnetic fields is an old problem of
quantum mechanics that plays a role in a wide variety of physical
problems, ranging from neutron stars \cite{shapiro} to the quantum
Hall effect \cite{prange}. From the point of view of quantum field
theory, several studies have been performed, related to the chiral
anomaly \cite{ishi}, effective actions \cite{cange,blau,dunne} and
applications to anyons systems \cite{wil}.

In this paper we would like to study another aspect of this
problem, namely, following the approach proposed in \cite{glv}, we
shall discuss the statistical mechanics of nonrelativistic
fermions embedded in a constant magnetic field.

Our approach requires the explicit computation of a fermionic
determinant, and this is carried out in terms of polylogarithm and
Dedekind functions. Additionally our calculation, in a particular
case, allows us to study the strong and weak magnetic field limits
quite easily, due the modular invariance of the Dedekind function.
The paper is organized as follows: in section 2 we review the
approach proposed in \cite{glv}, in section 3 we formulate the
problem in the context of the path integral method, in section
4 we present an explicit computation of the nonrelativistic grand
potential, and in section 5 we study the partition function in the
case of 2+1 dimensions. Our conclusions are given in section 6.

\section{Effective action at low energies}

In reference \cite{glv}, a method was proposed to compute
effective actions at low energies, based in a path integral
derivation of the Foldy-Wouthuysen transformation. This method can
be used for nonrelativistic fermions as well. Let us consider the
lagrangian
\begin{equation}
{\cal L}={\bar{\psi}}\bigl[{iD\hspace{-0.6em}\slash \hspace{0.15em}}-m\bigr]%
\psi ,  \label{laglh}
\end{equation}
where ${D\hspace{-0.6em}\slash \hspace{0.15em}}={\partial \hspace{-0.6em}%
\slash \hspace{0.15em}}+ig{A\hspace{-0.6em}\slash \hspace{0.15em}}$.

Following the procedures developed in \cite{glv}, we redefine the
origin of the energy by the rescaling
\begin{equation}
\psi (x)=e^{-imt}\phi (x),  \label{psi}
\end{equation}
to write the lagrangian as
\begin{equation}
{\cal L}={\bar{\phi}}(\,i{D\hspace{-0.6em}\slash \hspace{0.15em}}-m(1-\gamma
_{0}))\phi .  \label{lagheavy}
\end{equation}

Now we decompose the spinor $\phi$ into `large' ($\varphi$) and
`small' ($\chi$) components, in whose terms
\begin{eqnarray}
{\cal L} &=& {\varphi}^{\dagger} \,iD{_0} \varphi + {\chi}^{\dagger} \bigl[i
D{_0} + 2 m\bigr] \chi +{\varphi} ^{\dagger}\, i\,{\vec{ \sigma}}\cdot {\bf D%
} \chi  \nonumber \\
&&+{\chi}^{\dagger}\,i\, {\vec{\sigma}}\cdot {\bf D} \varphi,  \label{9}
\end{eqnarray}
where the Dirac representation for the $\gamma$-matrices has been
used.

The next step is to diagonalize the lagrangian in (\ref{9}) by
means of the change of variables
\begin{eqnarray}
\varphi^{^{\prime}} &=& \varphi,  \nonumber \\
{\varphi}^{^{\prime}\dagger} &=& {\varphi}^{\dagger},  \nonumber \\
{\chi}^{^{\prime}} &=& \chi + [ \,iD_0 + 2m]^{-1} \,i\,{{\vec{\sigma}}\cdot
{\bf D}} \varphi,  \label{11} \\
{\chi}^{^{\prime}\dagger} &=& {\chi}^{\dagger} +{\varphi}^{\dagger}\,i\,{{%
\vec{\sigma}} \cdot {\bf D}} \,[ \,iD_0 + 2m] ^{-1}.  \nonumber
\end{eqnarray}

This change of variables has a Jacobian equal to unity and the
effective lagrangian, under this transformations, becomes
(omitting the primes)
\begin{eqnarray}
{\cal L}&=&{\varphi}^{\dagger}\bigl[\,iD_0 + {{\vec{\sigma}}\cdot {\bf D}} \,%
{( \, iD_0 + 2m)} ^{-1}\,{{\vec {\sigma}}\cdot {\bf D}} \bigr] \,\varphi,
\nonumber \\
&+& \chi^{\dagger} \bigl[\, iD_0 + 2m \bigr] \chi.  \label{lfinal}
\end{eqnarray}

This lagrangian describes the non-local dynamics of the fermions
in terms of two-components spinors. One should note that $\varphi$
and $\chi$ decouple and, after expanding ${(\,iD_0 + 2m)} ^{-1}$
in  powers of $1/m$, the partition function becomes
\begin{equation}
Z = \int {\cal D} \chi^{\dagger} {\cal D } \chi e^{i S_{\chi}} \int {\cal D}
\varphi^{\dagger} {\cal D } \varphi e^{i S_{\varphi}},  \label{finpart}
\end{equation}
where
\begin{eqnarray}
S_{\chi} & = & \int d^4 x \chi^{\dagger} \bigl[\, iD_0 + 2m \bigr] \chi,
\nonumber \\
S_{\varphi} &=& \int d^4 x \varphi^{\dagger} \bigl [ \, i D_0 +\frac{1}{2m}
{\bf D}^2 + \frac{g}{2m} \vec{\sigma} \cdot {\bf B} \bigr] \varphi + O
(1/m^2).  \nonumber \\
& &  \label{exact}
\end{eqnarray}

$S_{\varphi}$ is the nonrelativistic action for fermions
interacting with a magnetic field. $S_{\chi}$ is the action
associated to the lower contribution of the spinor and can be
neglected in the nonrelativistic limit.

\section{Nonrelativistic fermions in magnetic fields}

Using (\ref{finpart}) and (\ref{exact}) one can explicitly study
the nonrelativistic quantum field theory of fermions at finite
temperature.

The partition function associated to this problem is
\begin{eqnarray}
Z_{\varphi } &=&\int {\cal D}\varphi {\cal D}\varphi ^{\dagger
}\,e^{iS_{\varphi }},  \nonumber \\
&=&\text{det }[iD_{0}+\frac{1}{2m}{\bf D}^{2}+\frac{g}{2m}\vec{\sigma}\cdot
{\bf B}+\mu],  \label{fer1} \\
&=&\prod_{n}\lambda _{n},  \label{fer2}
\end{eqnarray}
where $g$ stands for the electric charge of the fermions,
$\mu$ is the chemical potential,  and the infinite product
runs over all the eigenvalues $\lambda _{n}$ of the operator that
appears in (\ref{fer1}),
\begin{equation}
\lbrack iD_{0}+\frac{1}{2m}{\bf D}^{2}+\frac{g}{2m}\vec{\sigma}\cdot {\bf B} +\mu]\phi _{n}=\lambda _{n}\phi
_{n},  \label{fer3}
\end{equation}
subject to the usual antiperiodic boundary conditions in the
imaginary time direction.

For a constant magnetic field one can choose the gauge
\begin{eqnarray}
{\bf A}&=&(-B_{0}y,0,0), \nonumber
\\
A_0 &=& 0, \nonumber
\end{eqnarray}
in which case the fermionic determinant can be computed from
\begin{equation}
\lbrack i\partial _{t}+\frac{1}{2m}{\bf D}^{2}\pm \frac{g}{2m}B_{0}+ \mu]\phi
_{n}^{\pm }=\lambda _{n}^{\pm }\phi _{n}^{\pm },  \label{fer4}
\end{equation}
and thus the partition function (\ref{fer1}) becomes
\begin{eqnarray}
Z_{\varphi } &=&\text{det}[\,i\partial _{t}-H^{+}+\mu]\text{det}%
[\,i\partial _{t}-H^{-}+\mu],  \label{det1} \\
&=&\prod_{n}\lambda _{n}^{+}\prod_{n}\lambda _{n}^{-},  \label{det3}
\end{eqnarray}
where the $H^{\pm}$ can be read off from (\ref{fer4}).

Each determinant in (\ref{det1}) is evaluated by explicitly
solving
the eigenvalue equation (\ref{fer4}) by means of the
Ansatz $\phi _{n}^{\pm }({\bf x},t)=f_{n}^{\pm }({\bf x})T(t)$,
which yields
\begin{eqnarray}
\lbrack -\frac{1}{2m}{\bf D}^{2}\mp \frac{g}{2m}B_{0}-\mu]f_{n}^{\pm }
&=&(\Omega -\lambda _{n}^{\pm })f_{n}^{\pm },  \label{k1} \\
i{\dot{T}}(t)-\Omega T(t) &=&0,  \label{k2}
\end{eqnarray}
since the operator $\frac{1}{2m%
}{\bf D}^{2}\pm \frac{g}{2m}B_{0}$ is time-independent.
Equation (\ref{k1}) is just Schr\"{o}dinger's equation for the Landau
problem, whose eigenvalues are known,
\begin{equation}
E_{n}^{\pm }=\Omega -\lambda _{n}^{\pm }=(n+\frac{1}{2}\pm \frac{1}{2}%
)\omega +\frac{p_{z}^{2}}{2m} -\mu,  \label{k3}
\end{equation}
with $\omega =gB_{0}/m$. The equation for $T(t)$ has a solution
only if
\begin{equation}
\Omega=\Omega _{m}=\frac{\pi }{\cal T}(2m+1),  \nonumber
\end{equation}
where $\cal T$ is the period and $m$ an integer, in virtue of the
antiperiodic boundary condition on $T(t)$.

Thus, the eigenvalues in (\ref{det3}) are given by
\begin{equation}
\lambda_{m,n}^{\pm} = \frac{\pi}{\cal T} (2 m +1) - E_n^{\pm}.\label{au}
\end{equation}

The statistical mechanics of the fermionic system under
consideration is described by the grand potential --basically the
logarithm of the partition function after going to Euclidean
space, {\it i.e.} replacing $\cal T$ by $i\beta$, where
$\beta=1/T$ is the inverse temperature. The logarithm of the
partition function is
\begin{eqnarray}
\log Z_\varphi &=& \sum_{n=0}^{\infty} \sum_{m=-\infty}^{\infty}\int dp_z[\,\log(
\lambda_
{m,n}^+) +\log(\lambda_{m,n}^-)], \nonumber \\
&\equiv& \sum_{n=0}^{\infty} \int dp_z [ L_{n}^+ + L_n^-],  \label{logz}
\end{eqnarray}
where $L_n^{\pm }$ are infinite sums defined as
\begin{equation}
L^\pm_n = \sum_{m=-\infty}^\infty \log \,(\lambda^\pm_{m,n}). \label{def}
\end{equation}

Although the series in (\ref{def}) are divergent, they can be
computed by using a definite regularization prescription.

Let us start considering the divergent series
\begin{equation}
L(a,b)=\sum_{m=-\infty }^{\infty }\log \,(\,a\,m+b).
\label{k4}
\end{equation}
where $a$ and $b$ are constants.

The second derivative of $L(a,b)$ is
\begin{eqnarray}
\frac{d^{2}L(a,b)}{db^{2}} &=&-\sum_{m=-\infty }^{\infty
}\frac{1}{(am+b)^{2}},
\nonumber \\
&=&-\frac{\pi ^{2}\csc ^{2}(b\pi /a)}{a^{2}},  \label{k5}
\end{eqnarray}
so that upon integration one finds
\begin{equation}
L(a,b)=\log \bigg(e^{c_1 +b c_2 } \sin\bigg(\frac{\pi b}{a}\bigg)\bigg),
\end{equation}
where $c_{1}$ and $c_{2}$ are two arbitrary integration  constants.

In our case, $a$ and $b$ can be read from (\ref{au}) and
(\ref{def}). Therefore
\begin{eqnarray}
L^{\pm}_{n} &=& \log \bigg[e^{c_{1}^{\pm}+c_{2}^{\pm} [- i \frac{\pi
}{\beta} - E_{n}^{\pm }] }\cosh \bigg(\beta \frac{E_{n}^{\pm }}{2}\bigg) \bigg]
\label{k6}
\\
&\sim& c_{1}^{\pm} - i\frac{c_{2}^{\pm}\pi}{\beta} + (-c_{2}^{\pm} + \frac{\beta}{2})
E_{n}^{\pm} + \log( 1+ e^{-\beta E_{n}^{\pm }}). \label{def11}
\end{eqnarray}
Clearly, the constants  $c_{1}^{\pm}$ and $c_{2}^{\pm}$  parametrize the
arbitrariness of the regularization. Since the Euclidean space
effective action must be real, we choose $c^\pm_{1,2}$ in such a
way that (\ref{def11}) has no imaginary part.

At this point we can compare our result (\ref{def11}) with the
nonrelativistic limit of the general result given in \cite{cange}
for the effective action. The contribution
\[
c_{1}^{\pm} - i\frac{c_{2}^{\pm}\pi}{\beta} + (-c_{2}^{\pm} + \frac{\beta}{2})
E_{n}^{\pm}
\]
indeed corresponds -- after adding the analogous positron
contribution-- to the first term, ${\rm Tr}|{\cal E}|$, in equation
(18) of reference \cite{cange}, and the dependence on the
arbitrary constants $c_{1,2}^{\pm}$ reflects the regularization
that needs to be performed in order to define the trace.
For instance the choice $c_{1}^{\pm} = 0 = c_{2}^{\pm}$ is consistent with
$\zeta$-function regularization \cite{pepe}.

The second contribution, $\log( 1+ e^{-\beta E_{n}^{\pm }})$, which
is finite and independent of the constants $c_{1,2}^{\pm}$,
coincides (up to a factor $\beta$) with the grand potential in
nonrelativistic statistical mechanics.

\section{Grand-Potential}

In this section we compute the nonrelativistic grand potential
($\Omega$) directly from its thermodynamical definition, rather
than deriving it form the partition function ($Z$).

The nonrelativistic grand potential is given by
\begin{eqnarray}
\Omega(\beta,\mu)&=& -\frac{\tilde g}{\beta} \sum_{\ell=
+,-}\int_{-\infty}^{\infty} dp_z \sum_{n=0}^\infty  \log ( 1 + e^
{- \beta E_n^\ell} \, ),
\\
&=&\Omega^+(\beta,\mu) +\Omega^-(\beta,\mu),
\label{hurparti}
\end{eqnarray}
where  $\tilde g$ is the degeneracy factor $\tilde g =(g B_0 /4 \pi^2)  V$.

Clearly, we need to consider the following object
\ba
S(A,b)\equiv\sum_{n=0}^{\infty}\log\left(1+Ae^{-bn}\right).
\ea
In terms of $S$ we have
\ba
\Omega^{\pm}(T,\mu)=-\frac{\tilde{g}}{\beta}\int_{-\infty}^{\infty} \,dp_z\,S(A^{\pm},\beta
\omega)
\ea
with
\ba
A^{+}=e^{-\beta(\omega-\mu)}e^{-\beta p_z^2/2m},\quad
A^{-}=e^{\beta\mu}e^{-\beta p_z^2/2m},
\ea

The function $S(A,b)$ has the following Taylor series in the
variable $A$:
\ba
S(A,b)& = &\log (1 + A) + \sum\limits_{n = 1}^\infty  {\frac{{( - 1)^{n + 1} }}
{{n(e^{nb}  - 1)}}A^n },\\
&=&\sum\limits_{n = 1}^\infty  {\frac{{( - 1)^{n + 1} }}
{{n(1-e^{-nb})}}A^n }.
\ea
Since $A^n$ has the form $a^n e^{-n\gamma p_z^2}$, the $p_z$
integral can be performed explicitly to yield
\begin{equation}
\int_{ - \infty }^\infty  {dp_z S(A^ \pm  ,b) = \sqrt {\frac{{2\pi m}}
{\beta }} } \sum\limits_{n = 1}^\infty  {\frac{{( - 1)^{n + 1} }}
{{n^{3/2} (1 - e^{ - nb} )}}(a^ \pm  )^n }.
\label{o1}
\end{equation}
Expanding now
\[
(1 - e^{ - nb} )^{ - 1}  = \sum\limits_{k = 0}^\infty  {(e^{ - nb} )^k },
\]
we find
\begin{equation}
\int_{ - \infty }^\infty  {dp_z S(A^ \pm  ,b) =  - \sqrt {\frac{{2\pi m}}
{\beta }} } \sum\limits_{k = 0}^\infty  {\li_{3/2} ( - a^ \pm  e^{ - kb} )},
\label{o2}
\end{equation}
where we have introduced the polylogarithm function, defined as
the analytic continuation to the whole complex $z$-plane of the
series
\begin{equation}
\li_s (z) \equiv \sum\limits_{n = 1}^\infty  {\frac{{z^n }}
{{n^s }}},
\label{o3}
\end{equation}
defined for $|z| < 1$ and $s \in \Bbb C$.
Therefore,
\begin{equation}
\Omega ^ +  (T,\mu ) = V\frac{{gB_0}}
{{4\pi ^2 }} \frac{\sqrt{2\pi m}}
{{\beta ^{3/2} }} \sum\limits_{k = 0}^\infty  {\li
_{3/2} ( - e^{\beta \mu } e^{ - (k + 1)\beta \omega } )},
\label{o4}
\end{equation}
and
\begin{equation}
\Omega ^ -  (T,\mu ) = V\frac{{gB_0}}
{{4\pi ^2 }}\frac{\sqrt{2\pi m}}
{{\beta ^{3/2} }} \li_{3/2} ( - e^{\beta \mu } ) +
\Omega ^ +  (T,\mu ).
\label{o5}
\end{equation}

It is instructive to check that our result for
$\Omega=\Omega^++\Omega^-$ reduces to the correct result in the
limit of zero temperature ($\beta\to\infty$). Depending on the
sign of $\mu-k\omega$, the argument of the polylog function will
tend either to zero (in which case $\li_{3/2}(0)=0$ and there is
no contribution) or infinity. To get the corresponding
contribution in the latter case we need the asymptotic behavior of
$\li_{3/2}(z)$ as $z\to\infty$. This can be obtained from
Joncqui\`{e}re's relation \cite{emot},
\begin{equation}
\li_s (z) + e^{is\pi } \li_s (1/z) = \frac{{(2\pi )^s }}
{{\Gamma (s)}}e^{is\pi /2} \zeta (1 - s,\frac{{\log z}}
{{2\pi i}}).
\label{o6}
\end{equation}
In particular, for $x>0$ we have
\ba
\li_{3/2} ( - e^{\beta x} ) &\longrightarrow& \frac{{(2\pi )^{3/2} }}
{{\Gamma (3/2)}}e^{3i\pi /4} \zeta ( - \frac{1}
{2},\frac{1} {2} + \frac{{\beta x}} {{2\pi i}})\\
&\longrightarrow& - \frac{2} {3}\frac{{\beta ^{3/2} }} {{\Gamma
(3/2)}}x^{3/2},
\ea
because of the asymptotic behavior
$\zeta(-\frac{1}{2},q)\to-\frac{2}{3}q^{3/2}$ of the Hurwitz zeta
function. With $\Gamma(3/2)=\sqrt{\pi}/2$, we obtain the correct
nonrelativistic result
\begin{eqnarray}
\frac{1} {V}\Omega (T &=&0,\mu ) =  \nonumber
\\
&-& \frac{{2gB_0}} {{3\pi ^2
}}\sqrt {2m} \left[ {\sum\limits_{k = 0}^{\left\lfloor {\mu
/\omega } \right\rfloor } {(\mu  - k\omega )^{3/2}  - \frac{1}
{2}\mu ^{3/2} } } \right]. \label{final}
\end{eqnarray}
Here $\left\lfloor x\right\rfloor$ is the {\em floor} of the real
number $x$.

A similar calculation can be performed in 2+1-dimensions, but then the
result one obtains is nothing more than the starting expression
(\ref{hurparti}), with the corresponding 2+1-degeneracy factor
$\tilde{g}_2=(gB_0/2\pi)L^2$, $L^2$ being the area of the
2-dimensional quantization box:

\begin{eqnarray}
\Omega_{2+1}(T,\mu) = -\frac{\tilde{g}_2}{\beta}
 \Bigg[ 2&&\sum\limits_{k =
 0}^{\infty}\log\left(1+e^{\beta\mu}e^{-(k+1)\beta\omega}\right)
 \nonumber\\
 &&+\log\left(1+e^{\beta\mu}\right)\Bigg]. \label{palma}
\end{eqnarray}

\section{Partition Function in 2+1-Dimensions}

In section we compute the partition function in 2+1-dimensions.
In order to do that, one can proceed as follows:  the product (\ref{det3})
is first written as
\ba
Z_{\varphi}=\left(Z_0\right)^{\tilde{g}_2},
\ea
in view of the degeneracy $\tilde{g}_2$ of each Landau eigenvalue.
Using $\zeta$-function regularization the reduced partition
function $Z_0$ can be put into the form
\begin{equation}
Z_0 = \prod^{\infty}_{n=0} \cosh \bigg[\frac{E^+_n
\beta}{2}\bigg] \cosh \bigg[\frac{E^-_n \beta}{2} \bigg]. \label{zT}
\end{equation}

Using the explicit form of the eigenvalues for the Landau problem in
2+1 dimensions ($p_z=0$) from (\ref{k3}), (\ref{zT}) can be written as
\begin{equation}
Z_0 = \prod_{n=0}^{\infty }\cosh (\theta
n+\delta^{+})\cosh (\theta n+\delta^{-}).  \label{zT2}
\end{equation}
where
\begin{eqnarray}
\theta &=&\frac{\beta \omega}{2}, \nonumber \\
\delta ^{+} &=&\frac{\beta}{2} (\omega -\mu), \nonumber \\
\delta ^{-} &=&-\frac{\beta\mu}{2}.  \label{constant}
\end{eqnarray}

Defining $q= e^{-2\theta }$, $z=e^{-2\delta ^{+}}$ and
$\tilde{z}=e^{-2\delta ^{-}}$, then (\ref{zT2}) becomes
\begin{equation}
Z_0 = \prod_{n=0}^{\infty }e^{2n\theta }e^{\delta^++\delta^-} \left( 1+zq^
{n}\right)
\left( 1+\tilde{z}q^{n}\right).\label{zz}
\end{equation}

This is  most general expression for the (reduced) partition function in
2+1-dimensions.

A particular situation arises when the chemical
potential is zero. In such a case (\ref{zz}) becomes
\begin{equation}
Z_0= \left[\frac{\eta(2i\theta /\pi )}{\eta (i\theta
/\pi )}\right]^{2},  \label{zsimple}
\end{equation}
where
\begin{equation}
\eta (\tau )=q^{1/24}\prod_{n=1}^{\infty }(1-q^{n}),  \label{eta}
\end{equation}
is the Dedekind eta function \cite{handbook}, with $q=e^{2i\pi \tau
}$.

It is not difficult to see that formula (\ref{zsimple}) reproduces result
(\ref{palma}) in the case of vanishing chemical potential. Indeed, from
the definition of the eta function and writing
$1-q^{2n}=(1-q^n)(1+q^n)$ we find
\ba
Z_0=q^{1/12}\prod_{n=1}^{\infty }(1+q^{n})^2,  \label{eta2}
\ea
which, with $q=e^{-\beta\omega}$ leads to
\ba
\log
Z_{\varphi}=-\frac{1}{12}\beta\omega+2\sum_{n=1}^{\infty}
\log\left(1+e^{-n\beta\omega}\right).
\ea
This coincides with the result for $-\beta\Omega_{2+1}$ from
(\ref{palma}) at $\mu=0$, up to terms due to the regularization
used.
The high temperature limit (or equivalently the weak magnetic
field limit) corresponds to $q\to 1$, where the Dedekind function
can be approximated by
\cite{string}
\begin{equation}
\eta (\tau )\sim q^{1/24}(1-q)^{-1/2}e^{-\frac{\pi^{2}}{6(1-q)}}, \label{a0}
\end{equation}
and, therefore, the partition function in the high temperature
limit becomes
\begin{equation}
Z_0\big|_{\beta\omega\ll 1} \sim \frac{e^{\frac{5}{12} \beta \omega + \frac{2\pi^2}{3 \sinh \beta \omega}}}
{\cosh \frac{\beta \omega}{2}}. \label{hT}
\end{equation}

Moreover, as the Dedekind function satisfies the modular property
\begin{equation}
\eta (\tau ) =\bigg( \frac{i}{\tau }\bigg)^{1/2}\eta \bigg(- \frac{1}{\tau }\bigg),   \label{modular}
\end{equation}
one can compute the partition function in the strong
magnetic field limit (or low temperature limit) directly from (\ref{hT}).
Indeed, by using (\ref{hT}) and (\ref{modular}) we find
\begin{equation}
Z_0\big|_{\beta\omega\gg 1}  \sim  e^{-\frac{5\pi^2}{12\beta \omega}  -\frac{2\pi^2}
{3 \sinh \frac{2 \pi^2}{\beta \omega}}}
\cosh\biggl(\frac{\pi^2}{\beta \omega}\biggr). \label{lT}
\end{equation}

\section{Conclusions}

In this paper we have studied the motion of nonrelativistic
fermions in a constant magnetic field, following a quantum field
theory approach. The main point is the calculation of the
fermionic determinant by using a general method that contains all
the possible regularizations of the infinite product, in the form
of particular choices of the coefficients $c^\pm_{1,2}$.

In addition, we obtained explicit expressions for the
nonrelativistic grand-potential, in terms of a series of
polylogarithms, and for the partition function in 2+1 dimensions
and zero chemical potential, in terms of the Dedekind eta
function.
In the latter case, the modular properties of the Dedekind
function allow us to relate the strong and weak magnetic field
limits.
This relation could be useful in the dynamics of the quantum Hall effect.

\acknowledgments
We would like to thank G. V. Dunne by pointed out several
important references. This work was partially supported by grants
Fondecyt 8000017 (P.L.C.), 1010596, 2990037, 3000005 and
Dicyt-USACH.
\pagebreak


\begin{references}
\bibitem{shapiro} J. Shapiro and S. Teukolsky, Black Holes, {\it
White Dwarfs and Neutron Stars: The Physics of Compact Objects},
J. Wiley (1983).

\bibitem{prange}  R. B. Laughlin in  {\it The Quantum Hall Effect},
R. R. Prange and S. M Girvin (Eds.).

\bibitem{ishi}  K. Ishikawa, N. Maeda, cond-mat/ 0102347.

\bibitem{cange}  D. Cangemi and G. Dunne,   {\it Annals Phys.}
{\bf 249}, 582 (1996).

\bibitem{blau}  S. K. Blau, M. Visser and A. Wipf, {\it Int. J.
Mod. Phys.} {\bf A6}, 5409 (1991).

\bibitem{dunne} G.~Dunne and T.~M.~Hall, {\it Phys.\ Lett.\ } {\bf
B419}, 322 (1998).

\bibitem{wil} See {\it e.g.} F. Wilczek, {\it Fractional
Statistics and Anyon Superconductivity}, World Scientific ( 1987).

\bibitem{glv}  J. Gamboa, S. Lepe and L. Vergara. {\it Mod. Phys.
Lett.}  {\bf A} (in press), hep-ph/0007089.

\bibitem{landau}  L. Landau and E. Lifshitz, {\it Non-Relativistic
Quantum Mechanics}, Pergamon Press (1975).

\bibitem{pepe}  J. L. Cort\'{e}s, J. Gamboa, I. Schmidt and J.
Zanelli, {\it Phys. Lett.}  {\bf B444}, 451 (1998); J. L. Cort\'{e}s
and J. Gamboa, {\it Phys. Rev.}  {\bf D59} 105016 (1999).

\bibitem{olivier} C.~Dib and O.~Espinosa, {\it The magnetized
electron gas in terms of Hurwitz zeta functions}, math-ph/0012010,
to appear in {\it Nucl. Phys} {\bf B}.

\bibitem{handbook}  M. Abramowitz and I. Stegun Editors, {\it
Handbook of mathematical functions}, Dover Publications, 1970.

\bibitem{string}  See for example J. Polchinski, {\it String
Theory}, Vol 1, Cambridge University Press, 1998 or M. Green, J.
Schwartz and E. Witten, {\it Superstring Theory }, Vol 1,
Cambridge University Press, 1987.

\bibitem{landau} L. D. Landau and E. M. Lifshitz, Statistical
Physics, Pergamon Press (1975).

\bibitem{emot} A.~Erd\'{e}lyi, W.~Magnus, F.~Oberhettinger and
F.~Tricomi, {\it Higher transcendental functions}, Vol.1,
McGraw-Hill, 1953.

\end{references}
\end{document}